# Spatiotemporal vortex strings of light

Shunlin Huang[1], Ning Zhang[1,2], Xu Lu[1,2], Jun Liu[1,3,*], and Jinping Yao[1,*]

[1]*State Key Laboratory of High Field Laser Physics, Shanghai Institute of Optics and Fine Mechanics, Chinese Academy of Sciences, Shanghai 201800, China*

[2]*Center of Materials Science and Optoelectronics Engineering, University of Chinese Academy of Sciences, Beijing 100049, China*

[3]*Zhangjiang Laboratory, Shanghai 201210, China*

[*]Corresponding authors: jliu@siom.ac.cn; jinpingmrg@163.com

## Abstract

Light carrying orbital angular momentum (OAM) holds unique properties and boosts myriad applications in diverse fields from micro- to macro-world. Endeavors have been made to manipulate the OAM in order to generate on-demand structured light and to explore novel properties of light. However, the generation of an ultrafast wave packet carrying numerous vortices with various OAM modes, that is vortex string, has been rarely explored and remains a significant challenge. Moreover, methods that enable parallel detection of all vortices in a vortex string are lacking. Here, we demonstrate that a vortex string with 28 spatiotemporal optical vortices (STOVs) can be successfully generated in an ultrafast wave packet. All STOVs in the string can be randomly or orderly arranged. The diffraction rules of STOV strings are also revealed theoretically and experimentally. Following these rules, the topological charges and positions of all STOVs in a vortex string can be easily recognized. The strategy for parallel generation and detection of STOV strings will open up exciting perspectives in STOV-based optical communications and also promote promising applications of the structured light in light-matter interaction, quantum information processing, etc.

## Introduction

Lights with angular momentum, including orbital angular momentum (OAM) and spin angular momentum (SAM), have played an important role in a mass of fundamental and applied researches. Generally, SAM is related to polarization, while OAM is associated with spiral phase. Owing to the fact that the magnitude of OAM can be much larger than that of SAM, a light beam carrying OAM exhibits more intriguing properties. Since Allen et al. discovered that vortex beam can carry OAM[1], optical vortices have been extensively studied and applied in microparticle manipulation[2-4], microscopy[5], optical communications[6], and so on. The precise manipulation of OAM endows vortex beams with more degrees of freedom, which facilitates the discovery of new properties of light and opens up more widespread applications.

Vortex beams with spatial spiral phases usually carry longitudinal OAMs. The manipulation of OAM direction yields a novel spatiotemporal optical vortex (STOV) beam carrying a space-time spiral phase. Unlike conventional spatial vortex light, the STOV beam has a transverse OAM. It is only in recent years that STOV has been discovered experimentally[7], and controllably generated by using 4$f$ pulse shaper[8,9], although the transverse SAM has been demonstrated since ten years ago[10,11]. Recently, increasing attentions have been paid to explore basic properties of STOVs[12-30], such as the generation, propagation and diffraction[12-22], reflection and refraction[23], conservation of transverse OAM in harmonic generation[24-26], and spin-orbit interaction[27,28]. Moreover, detection technologies of STOV beams have been developed, such as interference method[8], transient-grating single-shot supercontinuum spectral interferometry[9], and diffraction method[14]. Despite these achievements, generation and detection of a wave packet carrying various STOVs remain elusive and need to be further explored.

A conventional vortex beam generally carries static OAM[1], that is the OAM does not change with time in a pulse. The manipulation of OAM in the time domain allows us to obtain dynamic or time-varying OAMs. Rego et al. successfully produced extreme-ultraviolet beams with time-varying OAMs via high harmonic generation[31]. Cruz-Delgado et al. synthesized a spatiotemporal wave packet with spectral and time-dependent OAM by using a two-stage reconfigurable module[32]. The generation of vortex light with dynamic OAMs results in the discovery of self-torqued property of light[31]. In the above-mentioned studies[31,32], the time-dependent OAM is longitudinal. Very recently, a wave packet carrying two first-order transverse OAM modes with opposite helicity was obtained at the position of tens of centimeters after the pulse shaper by using chirp pulses[33]. Note that if the two STOVs have the same helicity, it is difficult to distinguish the wave packet from a normal STOV with the topological charge (TC) $l = 2$, because it also has two phase singularities[8]. And the identification becomes more difficult if higher order STOV is embedded in a wave packet.

Although the time-varying longitudinal OAMs have been generated and the results are remarkable, the parallel generation of a STOV string with randomly arranged transverse OAMs has not been reported yet. Furthermore, the detection of vortex light with dynamic OAMs is usually complicated or time-consuming, because all OAM modes in a wave packet cannot be simultaneously recognized. Meanwhile, it is also a big challenge to distinguish a high-order STOV beam with static OAM from a wave packet carrying multiple STOVs with time-varying OAMs. Herein, we demonstrate that a wave packet carrying numerous STOVs with customizable TC arrangements can be controllably and flexibly generated in the far field. Furthermore, the TCs and positions of all STOVs in the wave packet can be recognized simultaneously using the diffraction method developed by us[14]. The generation of STOV strings will greatly extend the number of vortices with transverse OAMs in a single wave packet from single or few vortices to tens or even hundreds of vortices. The STOV string will open up a novel degree of freedom of vortex beams and inspire wide applications.

## Experimental setup

The experimental setup for generating and detecting STOV strings is depicted in Fig. 1. Laser pulses from a mode-locked laser are launched into a folded 4$f$ pulse shaper consisting of a grating, a cylindrical lens and a reflective spatial light modulator (SLM). The laser pulses are modulated in the space-frequency domain by the SLM placed in the Fourier plane. The phase pattern loaded onto the SLM is designed according to the TC arrangements of the STOV strings. The output pulses from the 4$f$ pulse shaper pass through a spherical lens, and then vortex strings carrying numerous STOVs are generated in the focal plane. The generated STOV strings are diffracted by the second grating. A cylindrical lens is placed after the second grating, and a charge-coupled device (CCD) is placed in the focal plane of the cylindrical lens to capture the diffraction patterns of the STOV strings. The diffraction pattern detection part can be regarded as a STOV-string detector. More experimental details are given in Methods.

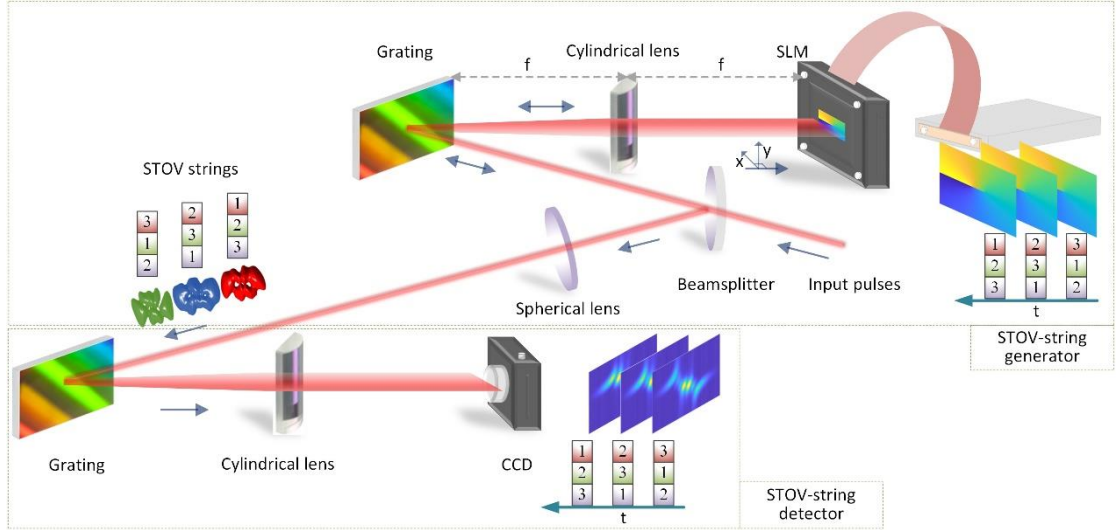

Fig. 1. Schematic of the experimental setup for generation and detection of STOV string with various STOVs. The STOV string is generated by a folded 4$f$ pulse shaper and a spherical lens. The pulse shaper consists of a grating, a cylindrical lens and a SLM. The configuration of the detection part is the same as that of the 4$f$ pulse shaper with the SLM replaced with a CCD.

To generate a vortex string with $n$ STOVs, a superposition phase, which is equivalent to the sum of $n$ spiral phases with offsets along the $x$ direction, is loaded onto the SLM. The phase is expressed as

$$\varphi = \sum_{j=1}^{n} l_j \cdot \tan^{-1}\left(\frac{y}{x+(j-1)\cdot d_x + \gamma_j}\right), \quad (1)$$

where $x$ and $y$ are the spatial coordinates perpendicular to the beam propagation direction. The dispersion direction of the grating is in the $x$ direction. $l_j$ is the TC value of the $j$th STOV. The distance between the $j$th STOV and the first STOV is $(j-1)\cdot d_x + \gamma_j$, where $\gamma_j$ can be used to adjust the gap between two neighboring STOVs. For STOVs with the same gap here, the STOV gap is $d_x$ and $\gamma_j = 0$.

## 3-STOV strings

As a proof-of-principle demonstration, a spatiotemporal wave packet carrying three STOVs, that is 3-STOV string, is generated by using an input pulse with a full width at half maximum (FWHM) spectral bandwidth of 6 nm. Here, the three STOV modes of $l$ = 1, 2, 3 are used, and the TC arrangements are selected to be 123, 231 and 312. The STOV string with the TC arrangement of −(231) is used for comparison. The calculated three-dimensional (3D) iso-intensity profiles and phase patterns of the 3-STOV strings are shown in rows 1 and 2 in Fig. 2, respectively. The wave packets of the 3-STOV strings no longer have elliptical structures compared with normal STOV with a single OAM state[8,13]. The wave packet shape as well as the positions and shapes of holes in the wave packet vary with the TC arrangement. All the spatiotemporal phases have six $2\pi$ phase windings, which results from that the total TCs of these strings are six. However, the phase patterns of these strings are different with each other owing to different TC arrangements. In addition, the wave packets (phases) of STOV strings with the same TC arrangement but contrary helicity (e.g., $l$:231 and $-l$:231) are mirror images of each other.

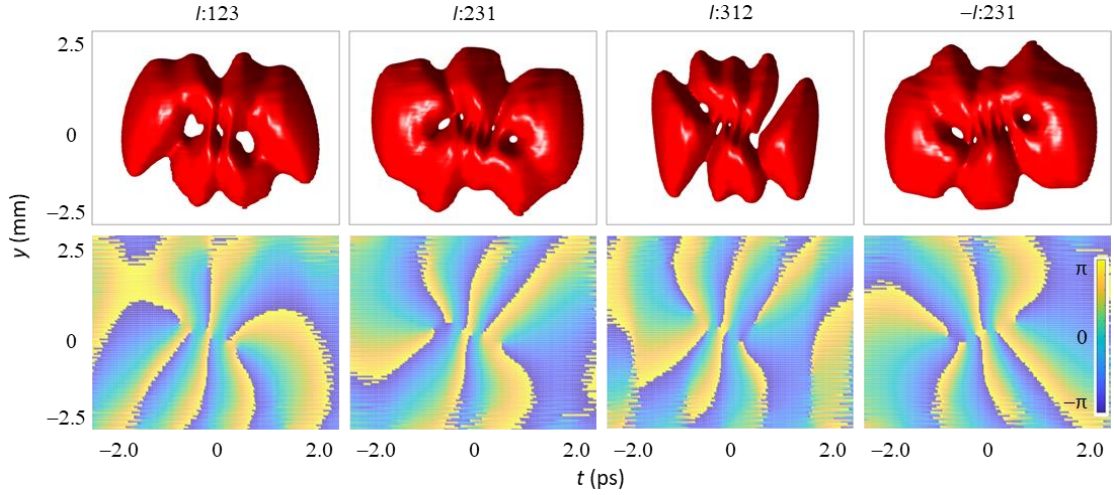

Fig. 2. The calculated intensity profiles and phase patterns of 3-STOV strings. Row 1 shows the isosurface plots of the 3-dimentional (*x*-*y*-*t*) wave packets generated in the far field, and they are viewed in the *t*-*y* plane. Rows 2 shows the corresponding phase patterns. The top row is marked as row 1.

As shown in Fig. 2, the 3-STOV string with the same helicity has multiple phase singularities in the wave packet, it is thus difficult to detect the TC of each STOV in the string by using the interference method[8] or transient-grating single-shot supercontinuum spectral interferometry[9]. Therefore, the diffraction method[14] is adopted here, which does not need time scanning, reference pulses, and complicated retrieved algorithm. The diffraction patterns of these STOV strings are shown in Fig. 3. Rows 1 and 2 show the isosurface plots of the simulated diffraction patterns and the corresponding intensity profiles. Row 3 shows the diffraction patterns captured directly by the CCD.

All diffraction patterns shown in Fig. 3 show multi-lobe structures with six gaps and seven lobes, which results from that the total TCs of these STOV strings are six. Each gap corresponds to 1 TC or $2\pi$ phase winding, which is the same as that of the normal STOV pulse carrying a single OAM mode with the same total TC[14]. The diffraction patterns have some unique features that can be used to identify different arrangements of STOVs in a wave packet and also to distinguish a STOV string from a normal high-order STOV with the same total TC. For a normal high-order STOV pulse, the diffraction pattern has two head lobes that are more energetic and larger than the inner lobes located between the two head lobes (see Fig. S3 of Supplement 1). Interestingly, as seen in Fig. 3, the diffraction patterns of STOV strings have more energetic lobes. Two lobes with higher energies are no longer located at two ends of the diffraction patterns, and their positions change with the TC arrangement. Each STOV in the string shares one end lobe with adjacent STOVs, and its diffraction pattern is the same as that of the corresponding STOV with a single OAM mode. Taking the STOV string with the TC arrangement of $l$:231 as an example, the first STOV with $l$ = 2 has 3 lobes, 2 energetic head lobes and 1 weaker inner lobe, which are marked as lobes 1, 2, 3. Similarly, lobes 3-6, lobes 6, 7 form the diffraction patterns of the second STOV with $l$ = 3 and the third STOV with $l$ = 1, respectively. For the STOV string, the shared lobes are lobe 3 and lobe 6, which have obviously higher energy than other inner lobes.

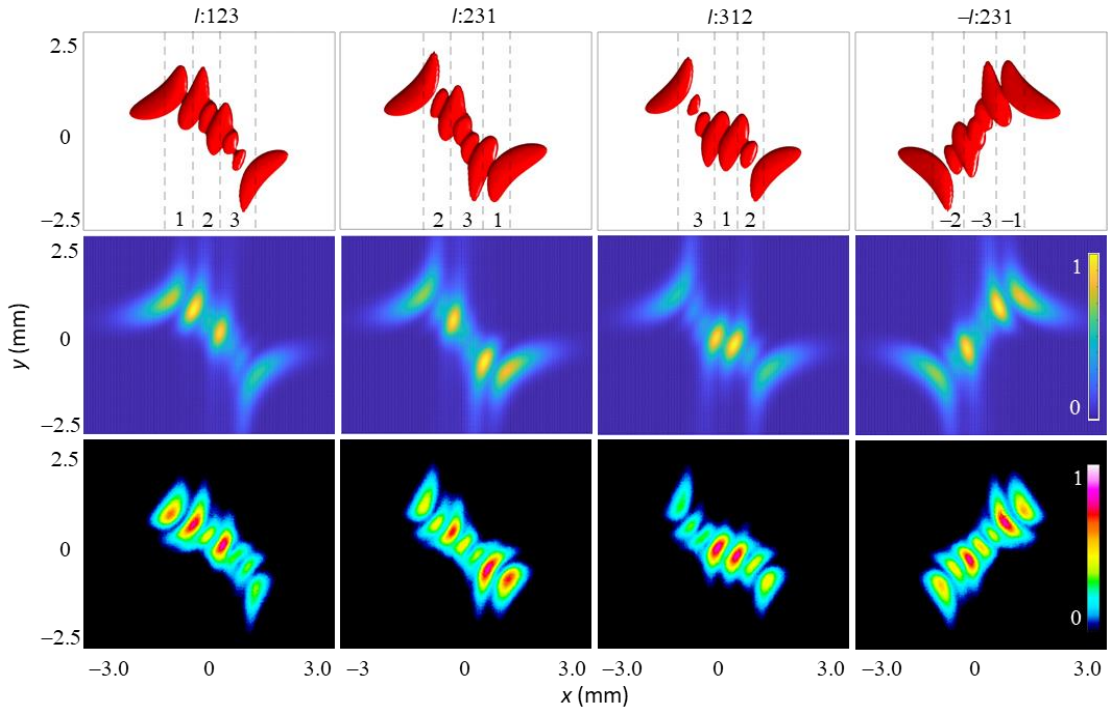

Fig. 3. Diffraction patterns of STOV strings with TC arrangements of $l$:123, $l$:231, $l$:312 and $-l$:231. Rows 1 and 2 show the simulated results. Row 1 shows the isosurface plots of the simulated diffraction patterns, which are viewed in the *x-y* plane. The diffraction patterns and TC values of three STOVs in the string are indicated with dashed lines and numbers, respectively. Row 2 shows the calculated intensity profiles of the diffraction patterns. Row 3 shows the measured results. The top row is marked as row 1.

For a vortex string with $m$ STOVs, there are $m-1$ shared lobes which have higher energy and larger size than the other lobes (excluding two end lobes). For 3-STOV strings shown with TC arrangements of $l$:123, $l$:231 and $l$:312 in Fig. 3, there are two shared lobes, which are lobes 2 and 4, lobes 3 and 6, lobes 4 and 5, respectively. Such kinds of diffraction rules have been well verified by the experimental results, as shown in row 3 in Fig. 3. Furthermore, the helicity of STOVs carried in a wave packet can be obtained from the orientation of diffraction pattern. For positive and negative TCs, the diffraction patterns are along the two diagonal directions. The unique spatial-resolved spectral intensity distribution in the diffraction pattern can be attributed to the transverse OAM-affected spectral intensity distribution (TOASID)[14]. From the characteristic diffraction pattern, the TC of each STOV and its position in the STOV string can be obtained simultaneously. Hence, the diffraction method enables the parallel detection of vortex string with numerous STOVs and the fast recognition of STOV strings with different TC arrangements. The 3-STOV string shown above can be regarded as one STOV-string cell, which facilitates the analysis of diffraction rules. Using the diffraction rules derived from the 3-STOV strings, longer STOV strings can also be recognized.

## 28-STOV strings

To improve STOV carrying capacity of STOV string in a wave packet, laser pulses with a broader spectral bandwidth (FWHM ~40 nm) are used. Such a broad spectrum allows us to generate vortex strings carrying 28 STOVs. The calculated intensity profile and phase pattern of STOV string carrying 28 STOVs with TC units of $-l$:123 are shown in Fig. 4a and b, respectively. The STOV gap is fixed at $d_x = 0.6$ mm. The wave packet of the long vortex string shows multi-slice structure, which is apparently different from the multi-hole structure of 3-STOV strings. The corresponding phase exhibits dramatic variations in the space-time domain, as shown in Fig. 4b. Apparently, it is hardly read out the TC arrangement information of the 28-STOV string from complex spatiotemporal wave packet.

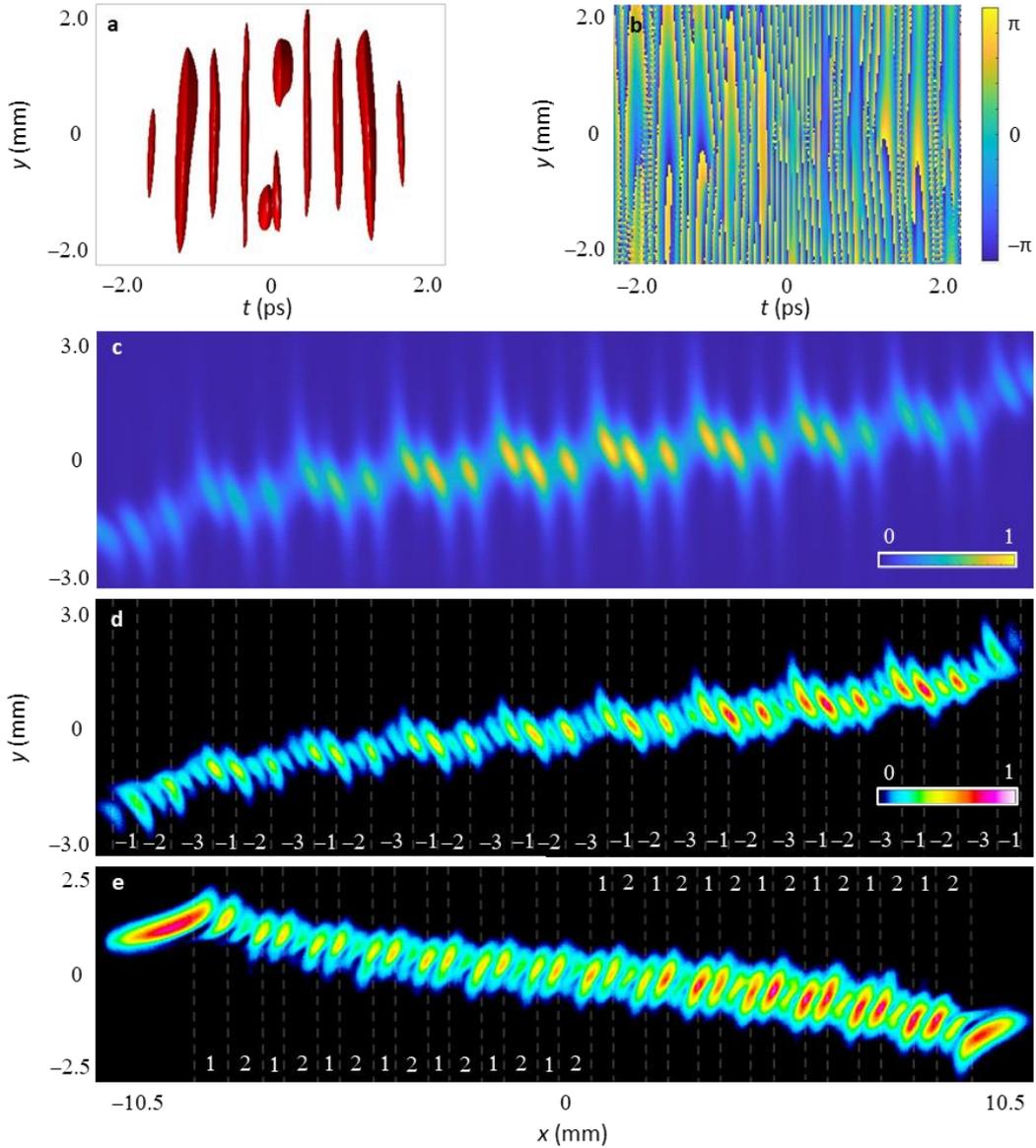

Fig. 4. The intensity profile, phase pattern, and diffraction patterns of 28-STOV string. a-d, The simulated 3D iso-intensity profiles, phase pattern, and the simulated and measured diffraction patterns of the 28-STOV string with TC unit of −$l$:123, respectively. e, The measured diffraction pattern of the 28-STOV string with TC unit of $l$:12.

Fortunately, the diffraction rules still apply for the vortex strings with numerous STOVs. Figure 4c and d shows the simulated and measured diffraction patterns of the wave packet in Fig. 4a, respectively. For clarity, the diffraction pattern of each STOV in the long string and the corresponding TC are marked with dashed lines and numbers, respectively. The measured diffraction pattern of the 28-STOV string matches with the simulated result very well. From the measured result, we can easily know the TC value, position and helicity of each STOV in the wave packet, all of which are key elements to characterize a specific STOV string. The diffraction pattern also clearly indicates that the 28-STOV string consists of a series of 3-STOV cells with the TC arrangement of $-l$:123. These results verify that our detection method is robust for both long and short STOV strings.

For comparison, Fig. 4e shows the measured diffraction pattern of the 28-STOV string with TC unit of $l$:12. Here, the STOV gap is $d_x = 0.5$ mm which is the same as the gap used in Fig. 3. In this case, the presence of two long end lobes indicates that the spectrum of input pulses is not fully utilized and more STOVs can be carried in the pulse. For STOV strings carrying low-order TCs, such as $l = 1$ or 2, clear diffraction patterns can be obtained using small STOV gaps. When the string involves higher-order STOVs, larger gaps are required to avoid the ambiguity of diffraction patterns (see Fig. S4 of Supplementary). There is a balance between the clarity of diffraction patterns and the STOV number carried in a wave packet. The dynamic variation of $d_x$ with the TC will be beneficial to improving the STOV carrying capacity, which can be realized by optimal design of the superposition phase.

The generation and detection methods above are also applicable for STOV strings with random TC arrangements, as illustrated in Fig. 5. The 28-STOV string with randomly arranged TCs $l = -1, -2, -3$ is generated with $d_x = 0.6$ mm. The simulated and measured diffraction patterns show good agreement. Following the same diffraction rules, the diffraction patterns can also be well identified. Interestingly, unlike the diffraction patterns of STOV strings with orderly TC arrangements that linearly arranges along the diagonal directions, the diffraction diagram of the STOV string with random TC arrangement shows a curved shape. Particularly, when three consecutive STOVs with the same TC of $l = 1$ appear in the string, the diffraction diagram appears a plateau. Furthermore, other STOV sequences with the same TC of $l = 3$ or $l = 2$ generated in the string can also be well recognized in the diffraction pattern. These results demonstrate that long STOV strings with random and orderly TC arrangements can be parallelly detected using the diffraction method.

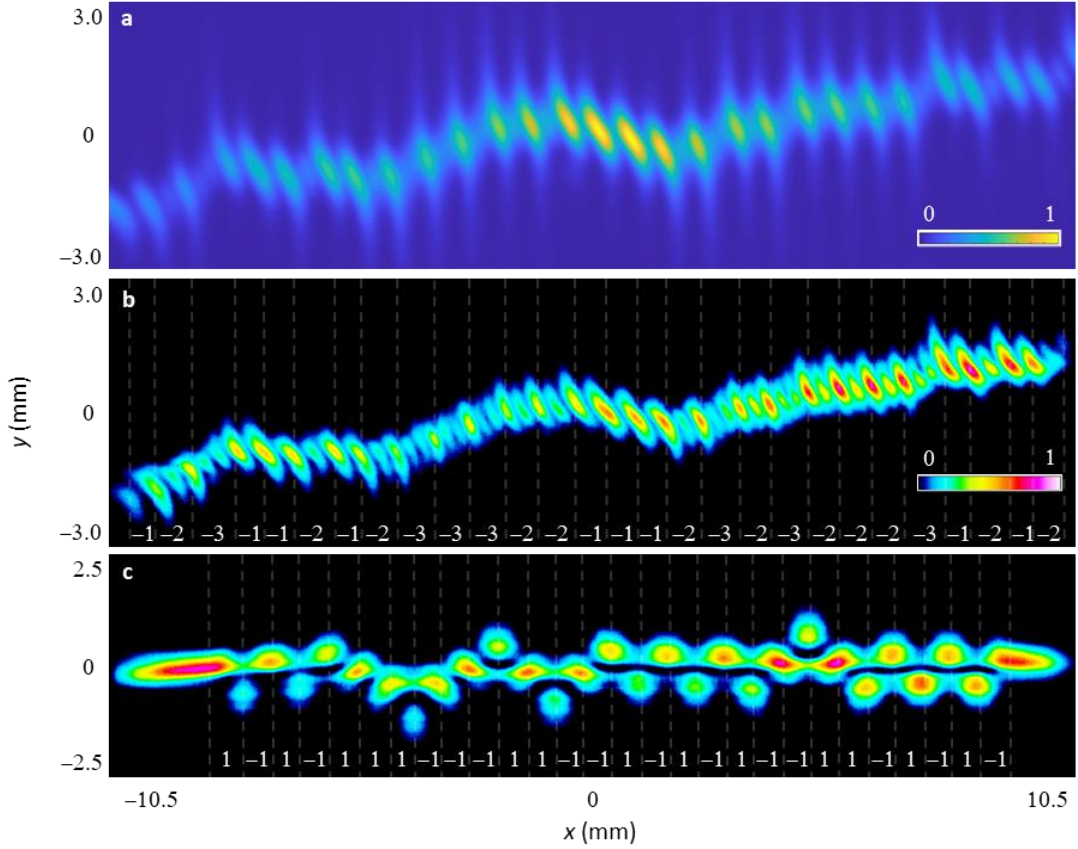

Fig. 5. Diffraction patterns of 28-STOV strings with random TC arrangements. a,b, Simulated and measured diffraction patterns of the STOV string carrying 28 STOVs with randomly arranged TCs $l = -1, -2, -3$, respectively. As it can be seen that, the simulated and measured results well match each other. c, The diffraction pattern of the STOV string carrying 28 STOVs with randomly arranged TCs $l = 1, -1$.

In addition, the STOV string simultaneously possessing positive and negative TCs can also be parallelly generated and detected. As an example, Fig. 5c shows the diffraction pattern of the 28-STOV string with randomly arranged TCs $l = 1, -1$. Unlike strip lobes in the diffraction patterns of STOV strings with the same helicity, the lobes shown here are close to circular. Moreover, the diffraction patterns of STOVs in the string are oriented along two diagonal directions, depending on the STOV helicity. This is actually a typical characteristic of STOV string carrying positive and negative TCs in a wave packet. The change in the arrangement orientation of these lobes indicates a flip in the helicity of two adjacent STOVs. Interestingly, the shared lobe between two

STOVs with contrary helicity is weaker than that with the same helicity. Furthermore, the lobe number shown in Fig. 5c cannot reflect the total TC of the STOV string. For a hybrid STOV strings carrying $m$ STOVs with TCs of $l = 1$ and $-1$, the total lobe number is $m+1$, and one gap still corresponds to 1 TC. Since there is no any inner weak lobe for the STOV with $l = 1$ or $-1$, it may be more convenient to identify the TC value. This kind of STOV string could facilitate applications of vortex light in high-speed optical communication.

## Discussion

We theoretically predicted and experimentally validated that long STOV strings with various TC arrangements can be generated in a wave packet by the phase manipulation of the input pulses. Such STOV string carries the time-varying OAMs, and it does not interact with external objects or fields after being generated. Therefore, the STOV string can be regarded as another type of self-torqued light with time-varying transverse OAMs compared with that with time-varying longitudinal OAMs[31].

Since the generation of STOV strings and the detection of TCs and positions of the STOVs can be considered as coding and decoding processes, respectively, our results would show promising applications in optical communication. A long STOV string enables us to realize high-bit coding/decoding merely utilizing several STOV modes. For instance, $m$-STOVs strings with 2 transverse OAM states can be used to achieve $m$-bit coding/decoding, which can be referred to as multi-state transverse OAM shift keying (MS-TOAMSK). The schematic of coding/decoding using transverse OAM is shown in Fig. S5 of Supplementary

In this article, the generated 28-STOV strings will be able to support 28-bit coding/decoding, which is difficult to accomplish for a spatial vortex beam by using OAM shift keying (OAMSK) as it requires $2^{28}$ OAM modes[34]. Furthermore, 28-STOV strings with $n$ different TCs ($n>2$) will have $n^{28}$ TC arrangements, which will have potential applications in a wide range of fields such as optical computing. By using larger SLMs and input pulses with a broader spectrum, more STOVs can be embedded in a wave packet. The generation of 100-STOVs strings may provide numerous opportunities in many areas of science and technology. It is noted that we just demonstrate the capability of MS-TOAMSK here, while a practical communication network is also related to many other conditions, such as the frame rate of SLM, which are beyond the scope of this work and will be studied in the future.

In summary, we demonstrated that numerous STOVs with various transverse OAMs can be carried in one wave packet. The vortex strings carrying 28 STOVs with orderly or randomly arranged TCs can be conveniently and flexibly generated in the far field without using predefined or special chirp pulses. Such long vortex strings can be parallelly recognized using the diffraction method. It is expected to produce vortex strings with 100 STOVs by increasing the size of SLMs and extending the spectral bandwidth of input pulses, which will pave the way for 100-bit coding/decoding in optical communication. In principle, the presented generation and detection methods can be applied to other spectral regions. This work will inspire widespread applications of vortex beams in nanophotonics, laser micromachining, quantum information processing, and light-matter interaction.

## Methods

**Experimental details.** The laser source is provided by a Ti:sapphire mode-locked laser (Micra, Coherent) with a repetition rate of 80 MHz and a central wavelength at 790 nm. A ruled reflective grating (1200 grooves/mm, blaze wavelength 750 nm) and a cylindrical lens ($f = 300$ mm) are applied in the folded $4f$ pulse shaper. An adjustable slit is placed after the cylindrical lens to tailor the spectral bandwidth and central wavelength of input pulses. In this experiment, the spectral width (FWHM) is chosen for ~6 nm or ~40 nm, and the central wavelength is fixed at 800 nm. A reflective SLM (P1920-0785-HDMI, Meadowlark Optics) with 1920×1152 pixels and a pixel size of 9.2 μm, which is placed in the Fourier plane of the pulse shaper, is used to modulate the input pulses. The phase pattern is designed according to the TC arrangement of the target STOV string and then is loaded onto the SLM. The modulated pulses are reflected to a spherical lens ($f = 1$ m) by a 50/50 beamsplitter. The target STOV string is produced in the focal plane of the spherical lens, i.e., far field. The second grating (1200 grooves/mm, blaze wavelength 750 nm) is placed in the focal plane of the spherical lens to diffract the wave packet of the generated STOV string. A cylindrical lens ($f$ = 300 mm) is placed 300 mm after the second grating. A CCD camera (WinCamD-LCM, DataRay) is placed in the focal plane of the cylindrical lens to capture the diffraction patterns of the STOV strings.

## Acknowledgements

This work is supported by the National Natural Science Foundation of China (Nos. 12034013, 12274428), Project of Chinese Academy of Sciences for Young Scientists in Basic Research (No. YSBR-042), Program of Shanghai Academic Research Leader (No. 20XD1424200), and Natural Science Foundation of Shanghai (Nos. 22ZR1481600, 23ZR1471700, 20ZR1464500).

## Author contributions

J.Y., J.L. and S.H. conceived the idea. S.H. performed the simulations and experiments. S.H., J.Y. and J.L. analyzed the data. S.H. and J.Y. prepared the manuscript and discussed with all authors.

## Competing interests

The authors declare no competing financial interests.

## References

1. Allen, L., Beijersbergen, M. W., Spreeuw, R. J. C. & Woerdman, J. P. Orbital angular momentum of light and the transformation of Laguerre-Gaussian laser modes. *Phys. Rev. A* **45**, 8185-8189 (1992).
2. Paterson, L. *et al.* Controlled rotation of optically trapped microscopic particles. *Science* **292**, 912-914 (2001).
3. He, H., Friese, M. E. J., Heckenberg, N. R. & Rubinsztein-Dunlop, H. Direct observation of transfer of angular momentum to absorptive particles from a laser beam with a phase singularity. *Phys. Rev. Lett.* **75**, 826-829 (1995).
4. Huang, S., Wang, P., Shen, X., Liu, J. & Li, R. Multicolor concentric ultrafast vortex beams with controllable orbital angular momentum. *Appl. Phys. Lett.* **120**, 061102 (2022).
5. Yan, L. *et al.* Q-plate enabled spectrally diverse orbital-angular-momentum conversion for stimulated emission depletion microscopy. *Optica* **2**, 900-903 (2015).
6. Wang, J. *et al.* Terabit free-space data transmission employing orbital angular momentum

multiplexing. *Nat. Photon.* **6**, 488-496 (2012).

7. Jhajj, N. *et al.* Spatiotemporal optical vortices. *Phys. Rev. X* **6**, 031037 (2016).

8. Chong, A., Wan, C., Chen, J. & Zhan, Q. Generation of spatiotemporal optical vortices with controllable transverse orbital angular momentum. *Nat. Photon.* **14**, 350-354 (2020).

9. Hancock, S. W., Zahedpour, S., Goffin, A. & Milchberg, H. M. Free-space propagation of spatiotemporal optical vortices. *Optica* **6**, 1547-1553 (2019).

10. Aiello, A., Lindlein, N., Marquardt, C. & Leuchs, G. Transverse angular momentum and geometric spin Hall effect of light. *Phys. Rev. Lett.* **103**, 100401 (2009).

11. Aiello, A., Banzer, P., Neugebauer, M. & Leuchs, G. From transverse angular momentum to photonic wheels. *Nat. Photon.* **9**, 789-795 (2015).

12. Hancock, S. W., Zahedpour, S. & Milchberg, H. M. Mode structure and orbital angular momentum of spatiotemporal optical vortex pulses. *Phys. Rev. Lett.* **127**, 193901 (2021).

13. Huang, S., Wang, P., Shen, X. & Liu, J. Properties of the generation and propagation of spatiotemporal optical vortices. *Opt. Express* **29**, 26995-27003 (2021).

14. Huang, S., Wang, P., Shen, X., Liu, J. & Li, R. Diffraction properties of light with transverse orbital angular momentum. *Optica* **9**, 469-472 (2022).

15. Cao, Q. *et al.* Sculpturing spatiotemporal wavepackets with chirped pulses. *Photonics Res.* **9**, 2261-2264 (2021).

16. Gui, G. *et al.* Single-Frame Characterization of Ultrafast Pulses with Spatiotemporal Orbital Angular Momentum. *ACS Photonics* **9**, 2802-2808 (2022).

17. Wan, C., Cao, Q., Chen, J., Chong, A. & Zhan, Q. Toroidal vortices of light. *Nat. Photon.* **16**, 519-522 (2022).

18. Wang, H., Guo, C., Jin, W., Song, A. Y. & Fan, S. Engineering arbitrarily oriented spatiotemporal optical vortices using transmission nodal lines. *Optica* **8**, 966-971 (2021).

19. Huang, J., Zhang, J., Zhu, T. & Ruan, Z. Spatiotemporal differentiators generating optical vortices with transverse orbital angular momentum and detecting sharp change of pulse envelope. *Laser Photon. Rev.* **16**, 2100357 (2022).

20. Wan, C., Chen, J., Chong, A. & Zhan, Q. Photonic orbital angular momentum with controllable orientation. *Natl. Sci. Rev.*, nwab149 (2021).

21. Chen, J., Wan, C., Chong, A. & Zhan, Q. Experimental demonstration of cylindrical vector spatiotemporal optical vortex. *Nanophotonics* **10**, 4489-4495 (2021).

22. Zang, Y., Mirando, A. & Chong, A. Spatiotemporal optical vortices with arbitrary orbital angular momentum orientation by astigmatic mode converters. *Nanophotonics* **11**, 745-752 (2022).

23. Mazanov, M., Sugic, D., Alonso, M. A., Nori, F. & Bliokh, K. Y. Transverse shifts and time delays of spatiotemporal vortex pulses reflected and refracted at a planar interface. *Nanophotonics* **11**, 737-744 (2022).

24. Hancock, S. W., Zahedpour, S. & Milchberg, H. M. Second-harmonic generation of spatiotemporal optical vortices and conservation of orbital angular momentum. *Optica* **8**, 594-597 (2021).

25. Gui, G., Brooks, N. J., Kapteyn, H. C., Murnane, M. M. & Liao, C.-T. Second-harmonic generation and the conservation of spatiotemporal orbital angular momentum of light. *Nat. Photon.* **15**, 608-613 (2021).

26. Fang, Y., Lu, S. & Liu, Y. Controlling photon transverse orbital angular momentum in high harmonic generation. *Phys. Rev. Lett.* **127**, 273901 (2021).

27   Chen, J., Yu, L., Wan, C. & Zhan, Q. Spin–orbit coupling within tightly focused circularly polarized spatiotemporal vortex wavepacket. *ACS Photonics* **9**, 793-799 (2022).

28   Bliokh, K. Y. Spatiotemporal vortex pulses: angular momenta and spin-orbit interaction. *Phys. Rev. Lett.* **126**, 243601 (2021).

29   Porras, M. A. Transverse orbital angular momentum of spatiotemporal optical vortices. Preprint at https://arxiv.org/abs/2301.09105 (2023).

30   Chen, W. *et al.* Time diffraction-free transverse orbital angular momentum beams. *Nat. Commun.* **13**, 4021 (2022).

31   Rego, L. *et al.* Generation of extreme-ultraviolet beams with time-varying orbital angular momentum. *Science* **364**, eaaw9486 (2019).

32   Cruz-Delgado, D. *et al.* Synthesis of ultrafast wavepackets with tailored spatiotemporal properties. *Nat. Photon.* **16**, 686-691 (2022).

33   Wan, C., Chen, J., Chong, A. & Zhan, Q. Generation of ultrafast spatiotemporal wave packet embedded with time-varying orbital angular momentum. *Sci. Bull.* **65**, 1334-1336 (2020).

34   Fu, S. *et al.* Experimental demonstration of free-space multi-state orbital angular momentum shift keying. *Opt. Express* **27**, 33111-33119 (2019).

# Supplementary information:

# Spatiotemporal vortex strings of light

## 1. Numerical simulations for generation and diffraction of STOV strings

A folded 4*f* pulse shaper with a SLM placed in the Fourier plane is used to generate STOV strings with various STOVs. The calculations for generation, propagation and diffraction of the STOV strings are the same as those used in previous works[S1,S2]. For simplicity, the laser pulses from a Ti:sapphire mode-locked laser are described as Gaussian pulses with Gaussian spatial profile and spectral distribution. They are used as the input light fields, and the pulse wave packet can be expressed as

$$E_1(x_1, y_1, \omega) = \exp[-(x_1^2 + y_1^2)/a^2]\exp[-(\omega-\omega_c)^2/b^2], \tag{S1}$$

where $\omega_c = 2\pi c/\lambda_c$ is the central frequency, $\lambda_c$ is the central wavelength, $c$ is the vacuum light velocity. $a$, $b$ are the waist radii of Gaussian profiles in the space and frequency domains, respectively. The input field is diffracted by the first grating, the first-order field immediately after the grating can be written as[S3]

$$E_2(x_2, y_2, \omega) = E_1(\beta x_2, y_2, \omega)\exp[i\gamma(\omega-\omega_c)x_2], \tag{S2}$$

where $\gamma = 2\pi/\omega_c d \cos(\theta_{d0})$, $\beta = \cos(\theta_{i0})/\cos(\theta_{d0})$, $\theta_{d0}$ and $\theta_{i0}$ are the diffracted angle and the incident angle of the central wavelength ray, respectively, and $d$ is the grating period. Then, the field $E_2(x_2, y_2, \omega)$ propagates to the first cylindrical lens. The field before the cylindrical lens can be calculated as[S4,S5]

$$E_3(x_3, y_3, \omega) = IFT\{FT\{E_2(x_2, y_2, \omega)\} \cdot H(f_x, f_y, \omega)\}, \tag{S3}$$

where *FT* and *IFT* denote the spatial Fourier transform and inverse Fourier transform, respectively. $H(f_x, f_y, \omega) = \exp[ikz - i\pi\lambda z(f_x^2 + f_y^2)]$, where $f_x$, $f_y$ are spatial frequencies in the *x* and *y* directions, respectively, and *k* is the wave vector. The wave packet after passing through the cylindrical lens can be written as[S6]

$$E_4(x_4, y_4, \omega) = E_3(x_3, y_3, \omega)\exp[-ikx_3^2/(2f)]. \tag{S4}$$

Then the light field continuously propagates to the Fourier plane. The field before the Fourier plane is marked as $E_5(x_5, y_5, \omega)$. Then the field is reflected by the SLM. The modulated field can be expressed as

$$E_6(x_6, y_6, \omega) = E_5(x_5, y_5, \omega)\exp(-i\varphi), \tag{S5}$$

where $\varphi$ is the superposition phase loaded onto the SLM.

The superposition phase loaded onto the SLM is equivalent to the sum of multiple spiral phase masks with transverse offsets along the $x$ direction, which is written as

$$\varphi = \sum_{j=1}^{n} l_j \cdot \tan^{-1}\left(\frac{y}{x+(j-1)\cdot d_x + \gamma_j}\right), \tag{S6}$$

where $x$ and $y$ are the spatial coordinates perpendicular to the beam propagation direction of the central ray. The dispersion direction of the grating is in the $x$ direction. $l_j$ is the topological charge (TC) of the $j$th STOV in the wave packet. The distance between the $j$th STOV and the first STOV is $(j-1)\cdot d_x + \gamma_j$, where $\gamma_j$ can be used to adjust the gap between two neighbor STOVs. When all the gaps are the same, $\gamma_j = 0$ and the gap is $d_x$.

The calculation of the light field after the reflection by the SLM is the same as discussed above. The STOV strings are obtained in the focal plane of a spherical lens (f = 1 m). The generated STOV strings are diffracted by the second grating placed in the focal plane of the spherical lens. The calculation of diffraction patterns of STOV strings is the same as the method used above with the input field replaced by the field of the STOV string.

The light field in the time domain is the inverse Fourier transform of the field in the frequency domain, then the intensity distribution of the generated STOV string in the space-time domain can be calculated as

$$I(x,y,t) = \left| \int_{-\infty}^{+\infty} E(x,y,\omega)\exp(i\omega t)d\omega \right|^2 . \tag{S7}$$

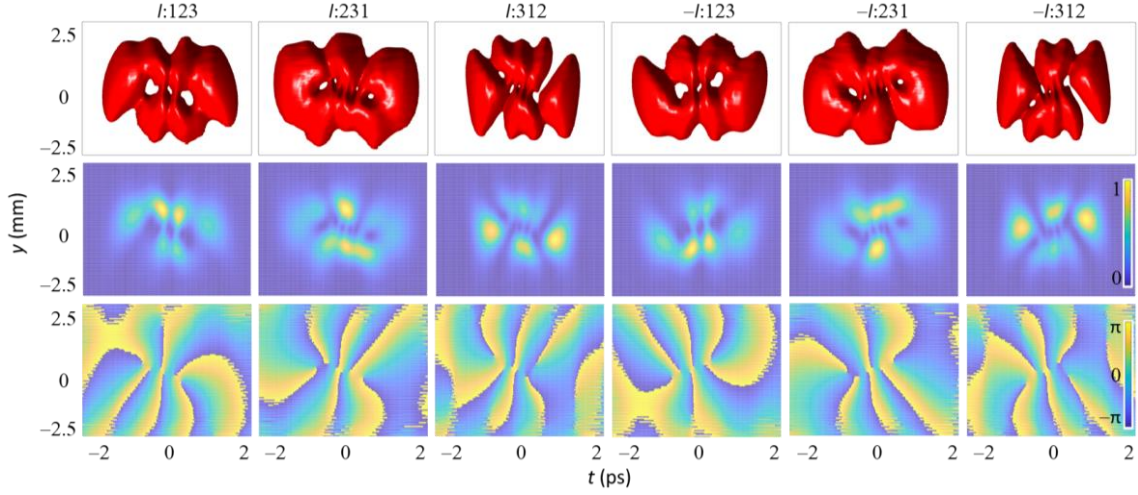

**Fig. S1** The calculated iso-intensity profiles and phase patterns of the 3-STOV strings. Row 1 shows the isosurface plots of the 3-dimentional (*x-y-t*) wave packets generated in the far field, and they are viewed in the *t-y* plane. Rows 2 and 3 show corresponding intensity profiles and phase patterns, respectively. The top row is marked as row 1.

The calculated results of the 3-STOV strings with the TC arrangements of 123, 231, 312, −(123), −(231), and −(312) are shown in Fig. S1. Row 1 shows the isosurface plots of the 3-dimentional (*x-y-t*) wave packets generated in the far field, and they are viewed in the *t-y* plane. Row 2 shows the corresponding intensity profiles. Row 3 shows their phase patterns. The wave packets of the 3-STOV strings are different from those of normal STOVs. The wave packets of normal STOVs show elliptical structures with multiple holes, and for high-order normal STOVs, such as $l$ = 6 STOV, the wave packets are split into two parts in the far field, as shown in the previous work[S2]. The space-time phases of all these STOV strings have six $2\pi$ phase windings, but the phase patterns are different from each other owing to different TC arrangements. Therefore, these wave packets show different shapes in the space-time domain, although they have the same total TC. The diffraction patterns of these STOV strings are shown in Fig. S2. The experimental results are in good agreement with the simulated results. From the diffraction patterns, the TCs, positions and helicities of all STOVs in a vortex string can be recognized.

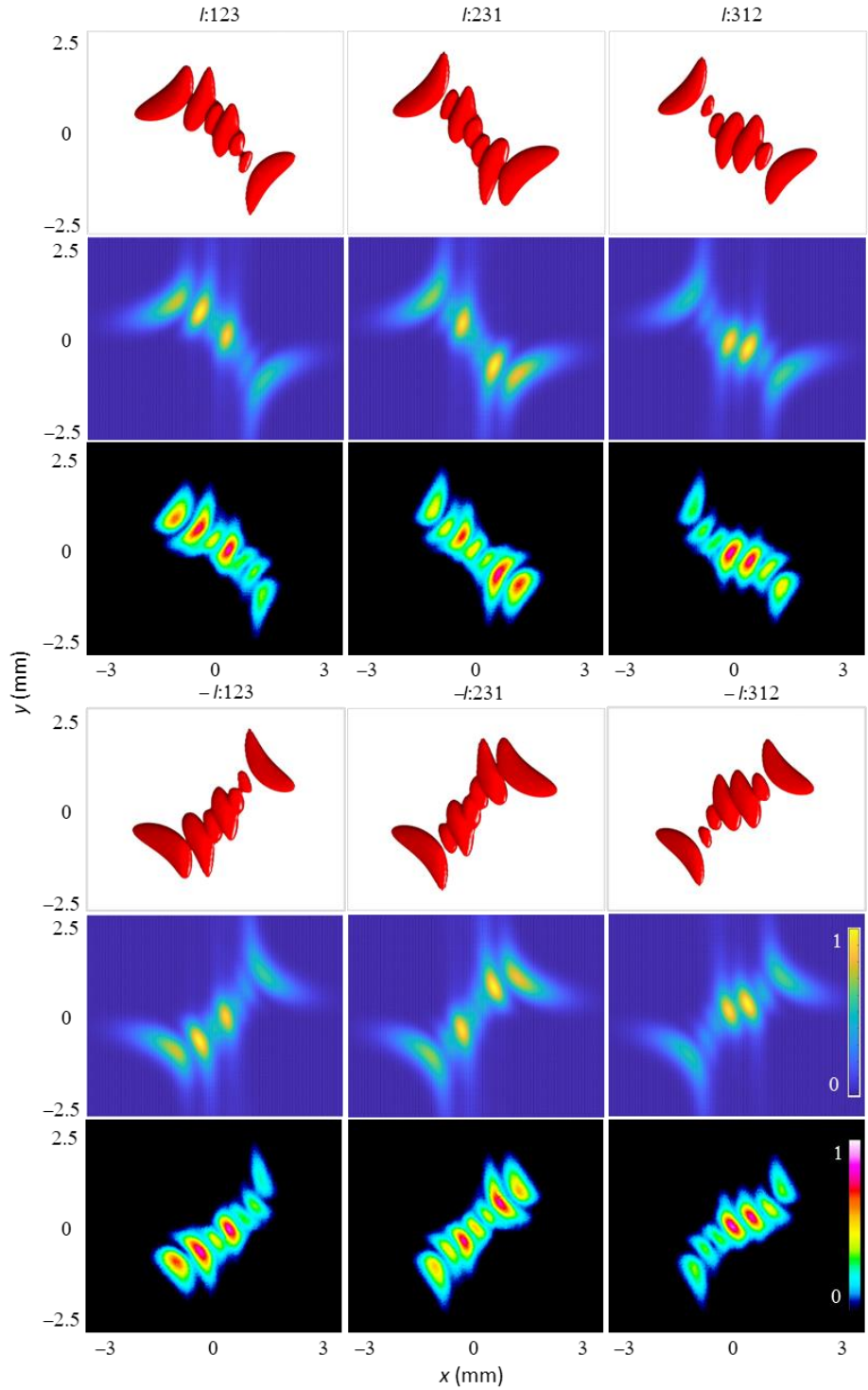

**Fig. S2** Diffraction patterns of STOV strings with TC arrangements of *l*:123, *l*:231, *l*:312, −*l*:123, −*l*:231, and −*l*:312. Rows 1 and 2 show the simulated results. Row 1 shows the isosurface plots of the simulated diffraction patterns, which are viewed in the *x-y* plane. Row 2 shows the calculated intensity profiles of the diffraction patterns. Row 3 shows the measured diffraction patterns by the CCD.

## 2. The diffraction patterns of normal STOVs with $l = \pm 6$

The measured diffraction patterns of normal STOVs with TCs of $l = \pm 6$ are shown in Fig. S3. The spectral bandwidths of the $l = \pm 6$ STOVs here are the same as those of the 3-STOV strings in Fig. S2. The diffraction patterns of the sixth order STOVs have 7 lobes with 6 gaps, and the two head lobes have higher energies and larger size than the inner lobes that located between the two end lobes, which are different from the diffraction patterns of the 3-STOV strings shown in Fig. S2. Therefore, we can well distinguish the STOV string from a normal high-order STOV when the total TC of the former is identical with that of the latter.

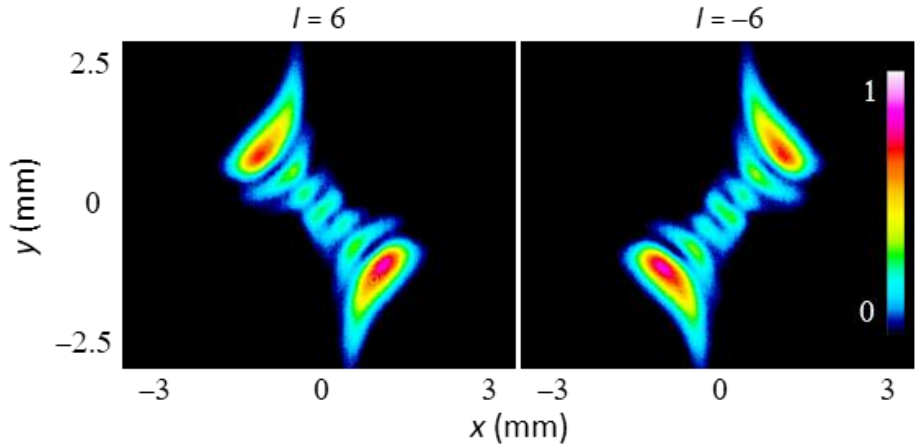

**Fig. S3** Measured diffraction patterns of normal $l = \pm 6$ STOVs.

## 3. The diffraction patterns of 20-STOV strings with TC units of $\pm l$:1234

The diffraction patterns of 20-STOV strings with TC units of $\pm l$:1234 are shown in Fig. S4. Clear diffraction patterns are demonstrated using a STOV gap of $d_x = 0.7$ mm. For each diffraction pattern, there are 51 lobes and 50 gaps, which results from that the total TC of the 20-STOV string is 50. As it can be seen, the high-order STOV with $l = \pm 4$ can still be identified in the 20-STOV strings when the STOV gap is increased to 0.7 mm. The diffraction rules and detection methods derived from the 3-STOV strings can also be used here.

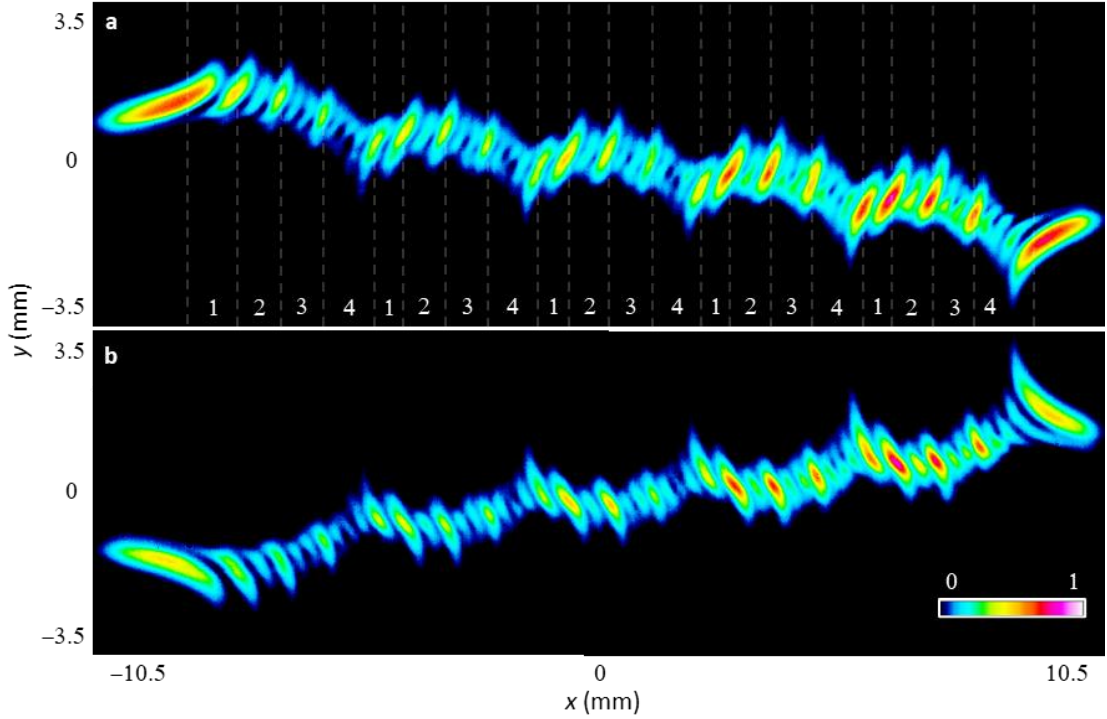

**Fig. S4** The measured diffraction patterns of 20-STOV strings with TC units of (a) $l$:1234 and (b) $-l$:1234.

## 4. The multi-state transverse OAM shift keying (MS-TOAMSK)

The generation of STOV strings and the detection of TCs and positions of the STOVs can be considered as coding and decoding processes, respectively. Thus, the generated STOV strings may be utilized in optical communication. Since a wave packet can carry numerous STOVs, we can realize high-bit coding/decoding merely using several STOV modes. For instance, for conventional vortex beam, one pulse only carries one vortex, thus $m$ OAM states are required to achieve a $\log_2 m$ bits coding/decoding, where $m$ OAM states act as $m$-ary symbols to indicate $m$-ary numbers, that is OAM shift keying (OAMSK) [S7]. However, the pulse trains with each pulse carrying $m$ STOVs with two transverse OAM states can be used to achieve $m$-bit coding/decoding, where the positions of the STOVs in the pulse can be utilized as an additional dimension. It thus can be referred to as multi-state transverse OAM shift keying (MS-TOAMSK). The schematic of coding/decoding using transverse OAM in data transmission is shown in Fig. S5. In this work, 28 STOVs can be generated in one wave packet, which would support 28-bit coding/decoding. It is difficult to achieve for conventional vortex beams by using OAMSK where $2^{28}$ OAM modes are required.

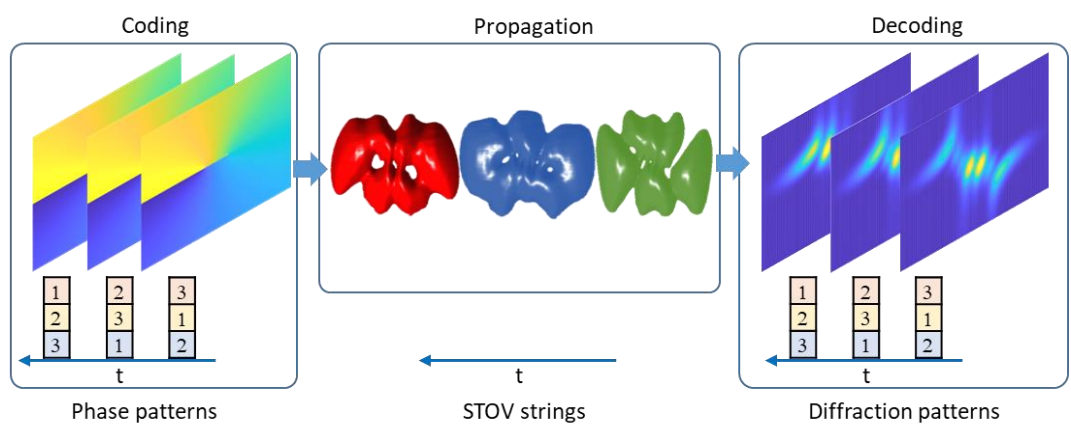

**Fig. S5** The schematic of data transmission using STOV strings, namely multi-state transverse OAM shift keying (MS-TOAMSK). The three colored wave packets shown in the middle represent three 3-STOV strings with TC arrangements of 123, 231 and 312.

By using pulses with broader spectrum and larger SLMs, we can produce vortex strings with more STOVs in a wave packet. The 100-bit coding/decoding is expected to realize, which wound have application in high-speed optical communications. Moreover, the combination of MS-TOAMSK with other multiplexing methods such as polarization multiplexing would further promote communication speed and capacity. Note that, here, we just demonstrate the capability of MS-TOAMSK, while a practical communication network may be related to many other conditions, such as the frame rate of SLM, which are beyond the scope of this work and will be studied in our future work.

## References

S1  Huang, S., Wang, P., Shen, X., Liu, J. & Li, R. Diffraction properties of light with transverse orbital angular momentum. *Optica* **9**, 469-472 (2022).
S2  Huang, S., Wang, P., Shen, X. & Liu, J. Properties of the generation and propagation of spatiotemporal optical vortices. *Opt. Express* **29**, 26995-27003 (2021).

S3  Martinez, O. E. Grating and prism compressors in the case of finite beam size. *J. Opt. Soc. Am. B* **3**, 929-934 (1986).

S4  Goodman, J. W. *Introduction to Fourier Optics*. Roberts and Company Publishers, (2005).

S5  Ratcliffe, J. A. Some aspects of diffraction theory and their application to the ionosphere *Rep. Prog. Phys.* **19**, 188-267 (1956).

S6  Denisenko, V. *et al.* Determination of topological charges of polychromatic optical vortices. *Opt. Express* **17**, 23374-23379 (2009).

S7  Fu, S. *et al.* Experimental demonstration of free-space multi-state orbital angular momentum shift keying. *Opt. Express* **27**, 33111-33119 (2019).